\documentclass{iopconfser}
\usepackage{graphicx}
\usepackage{mathtools}

\begin{document}

\title{An explicit statistical model for the Bell experiment}

\author{David H. Oaknin$^{1}$}

\affil{$^1$Rafael Ltd, IL-31021 Haifa, Israel}

\email{d1306av@gmail.com}

\begin{abstract}
Solid experimental evidence has now been obtained that confirms the violation of Bell's inequality in tests of maximally entangled qubit pairs. This violation is widely interpreted as definitive proof of the impossibility of describing quantum phenomena in terms of locally defined elements of reality. In a series of recent papers, we have noticed, however, that this conclusion inadvertently, yet crucially, relies on the assumed existence of an absolute frame of reference, with respect to which it would be possible to describe independently of each other the hypothetical elements of reality and the measurement devices that test them. Otherwise, a non-zero geometric phase may appear in the description of the former with respect to a closed sequence of settings of the latter, leading to the violation of the inequality. Following this observation, we discuss an explicit statistical model, which fully reproduces the predictions of Quantum Mechanics for the Bell experiment.  
\end{abstract}

\section{Introduction}
\vspace{0.2in}
 
The inability to complete the description of quantum systems provided by Scrodinger's wavefunction in terms of {\it elements of physical reality} as envisioned by Einstein, Podolsky and Rosen \cite{EPR} lies at the core of a long-lasting debate about the foundations of quantum theory and the role played by measurements. In the last sixty years, the debate has focused on several fundamental mathematical theorems that show that specific quantum mechanical predictions cannot be reproduced in terms of such elements of reality within any theoretical framework that shares certain intuitive features \cite{Bell, K-S, CHSH, CH, GHZ, Leggett, Colbeck, Cabello1,Fine,Hall}. 

The best known among these theorems is Bell's theorem \cite{Bell, CHSH, Fine}, which puts constraints (Bell's inequalities) on the correlations between the outcomes of measurements within the framework of these generic models. These constraints allow us to experimentally distinguish between the predictions of Quantum Mechanics, which violate them, and those of these generic models, which must abide by them. 

Carefully designed experiments performed during the last forty years have consistently confirmed the violation of Bell's inequalities according to the predictions of Quantum Mechanics \cite{Aspect, Tittel, Weihs, Matsukevich, Rowe, Zeilinger, Branciard, Cabello2, Cabello3, Hansen}. This solid experimental evidence is widely interpreted (see, for example, \cite{Wiseman}) as definitive proof of the impossibility of describing Bell's experiment in terms of the elements of reality envisioned by Einstein, Podolsky and Rosen \cite{EPR}.  

In a series of recent papers \cite{OakninFrontiers,OakninModPhys,OakninKalevHen}, we have noticed, however, that Bell's theorem relies crucially on an implicit assumption whose fulfillment is not guaranteed by fundamental physical principles. Namely, the existence of an absolute frame of reference with respect to which the hypothetical elements of reality and the settings of the measurement devices that test them can be described independently of each other. In the absence of such a frame, the theorem does not hold due to the possible appearance of a non-zero geometric phase in the description of the elements of reality with respect to a closed sequence of settings of the devices, and Bell's inequality can be violated. A geometric phase is allowed because the relative orientation of the two measurement devices entirely determines their setting since a rigid rotation of the two corresponds to a nonphysical gauge degree of freedom that must not play any role in the description of the system.

The paper is organized as follows. In section 2, we review the notion of geometric phases in gauge theory. In Section 3 we describe in detail how this well-known physical concept might offer a physically intuitive solution to the EPR paradox, and present an explicit statistical model that fully reproduces the predictions of Quantum Mechanics for the Bell experiment. We summarize our conclusions in section 4. 

\section{Geometric phases in gauge theories}
\vspace{0.2in}

As beautifully described in \cite{Wilczek, Wilczek1}, tiny bacteria self-propel within their environment while keeping at all time zero total momentum, which might seems paradoxical. At low Reynolds number, bacteria cannot exploit inertia to swim since the total forces acting on them by their environment are always strictly zero. Therefore, in order to propel, they perform cyclic sequences of body contortions that allow them to advance by acquiring in each cycle a non-zero geometric phase in the gauge degree of freedom that describes their location. Quite similarly, falling cats perform cyclic contortions that allow them to rotate their bodies and land on their legs by acquiring a rotational geometric phase while keeping a zero net angular momentum at all times \cite{Focus}.

The ultimate reason why a non-zero geometric phase may appear at the end of these closed cycles of body contortions is that the resulting rigid body rotations or translations correspond to gauge degrees of freedom. In this paper, we explore the possibility that a non-zero geometric phase associated with a gauge degree of freedom may be needed in theoretical models of the Bell experiment built in terms of local elements of reality. We will show that the Bell inequality does not constrain such models and, therefore, they are not ruled out by the collected experimental evidence. In fact, we present in this paper an explicit model of this kind that fully reproduces the predictions of Quantum Mechanics for the Bell experiment \cite{OakninFrontiers, OakninModPhys, OakninKalevHen}. 

\section{An explicit statistical model for the Bell experiment}
\vspace{0.2in}

In a Bell experiment, two parties located at distant labs share pairs of maximally entangled qubits produced by a source, with one qubit of each pair going to each party. Upon arrival at their respective labs, the qubits' polarization is tested with the help of measurement devices whose orientations can be freely and independently set. The outcome of each measurement is a binary variable taking values in the domain $\left\{-1,+1\right\}$. At any setting of the detectors, the average outcome of each one of them is zero while the correlation between their outcomes is a function of their relative orientation ${\widetilde \Delta}$ given by,

\begin{equation}
E({\widetilde \Delta}) = -\cos({\widetilde \Delta}).    
\end{equation}
\\
These correlations violate Bell's inequality \cite{Bell}, and their experimental confirmation \cite{Aspect, Tittel, Weihs, Matsukevich, Rowe, Zeilinger, Branciard, Cabello2, Cabello3, Hansen} is widely interpreted as a definitive proof of the impossibility of describing quantum phenomena in terms of the locally defined elements of reality envisioned by Einstein, Podolsky and Rosen \cite{Wiseman}. 

In this paper, however, we lay down an explicit theoretical framework that fully reproduces these predictions in terms of such elements. The crux of the model is the observation that the experimental setting of the measurement devices in a Bell experiment is fully characterized by their relative orientation ${\widetilde \Delta}$, while a rigid rotation of the two corresponds to a nonphysical gauge degree of freedom that must play no role in the description of the experiment. This observation is crucial because, as noticed above, gauge degrees of freedom may acquire a non-zero geometric phase over a closed sequence of modifications of the setting of the devices so that the requirements needed for Bell's theorem to hold are not fulfilled. 

Let us start by considering the two measurement devices as reference frames in which the respective incoming qubits and the binary responses they produce are naturally defined. For the sake of concreteness, we shall assume that each qubit is described with respect to its corresponding detector by an angular coordinate, which we denote respectively as 
\begin{equation}
\omega_A, \omega_B \in [-\pi, \pi),
\end{equation}
\\
while the responses they produce in their respective detectors are given by
\\
\begin{equation}
 S_A = S(\omega_A), \ \hspace{0.7in} \ S_B = S(\omega_B).   
\end{equation}
\\
where
\begin{equation}
\label{partition_1A}
S(\omega) =  \left\{
\begin{array}{ccccc}
+1,& \hspace{0.2in} \mbox{if} \hspace{0.2in} \omega \in & [& 0,& \pi), \\
-1,&  \hspace{0.2in} \mbox{if} \hspace{0.2in} \omega \in & [&-\pi,& 0),
\end{array}
\right.
\end{equation}  
These responses are explicitly locally determined since they depend only on how the incoming qubits appear with respect to the frame of reference set by their respective detectors. 
\\

Since the two qubits of each pair are twin produced together by the source, it is expected that their orientations with respect to their respective detectors will be related to each other by some strictly monotonic function that depends parametrically on the relative orientation between the two detectors,

\begin{equation}
\label{transformation}
\omega_B = -L(\omega_A ; {\widetilde \Delta}).
\end{equation}
\\
This relationship can be understood as a manifestation of a relativity principle that states how twin objects appear with respect to two different frames of reference.\\
\\

Since this relationship (\ref{transformation}) must not involve spurious gauge degrees of freedom, the relative orientation between the detectors is defined as

\begin{equation}
{\widetilde \Delta} = \Delta + \Phi,
\end{equation}
where $\Delta$ denotes a relative rotation of the detectors with respect to a previous reference setting characterized by a relative angle $\Phi$. 
This definition guarantees that subsequent relative rotations of the two detectors are additive, as demanded by experimental evidence. That is, two subsequent relative rotations of the detectors by amounts $\Delta_1$ and then $\Delta_2$ must result in a final rotation by a total amount $\Delta_1+\Delta_2$. In order to prove this feature of the model, let us start by considering a reference setting of the two detectors characterized by a reference angle $\Phi$. For this setting, the angular orientations of the pair of twin qubits with respect to their respective detectors are related by the transformation \\
\begin{equation}
\omega_B = -L(\omega_A; \Phi).    
\end{equation}
\\
After performing a relative rotation of the two detectors by an amount $\Delta_1$ with respect to the previous setting, the two angular coordinates are related by the transformation \\
\begin{equation}
\omega_B = -L(\omega_A; \Delta_1 + \Phi).    
\end{equation}
\\
Thus, this new setting is characterized by an angle $\Phi'=\Delta_1+\Phi$ so that after performing a second relative rotation of the two detectors by an angle $\Delta_2$, the two angular coordinates are related by \\
\begin{equation}
\omega_B = -L(\omega_A; \Delta_2 + \Delta_1 + \Phi).    
\end{equation}
\\
As we advanced, this corresponds to a total rotation by an amount $\Delta_1+\Delta_2$ with respect to the original reference setting characterized by an angle $\Phi$. \\
\\

Let us now consider the particular case in which the relationship (\ref{transformation}) is given as (see Fig. 1): 

\begin{itemize}
\item If  ${\widetilde \Delta} \in [0, \pi)$, 
\begin{eqnarray}
\label{Oaknin_transformation}
\hspace{-0.15in}
L(\omega; {\widetilde \Delta}) =  
\left\{
\begin{array}{c}
\hspace{-0.2in} q(\omega) \cdot \mbox{acos}\left(-\cos({\widetilde \Delta}) - \cos(\omega) - 1 \right), \\ \hspace{0.88in} \mbox{if}  \hspace{0.1in} -\pi \hspace{0.16in} \le  \omega < {\widetilde \Delta}-\pi, \\
\hspace{-0.2in} q(\omega) \cdot \mbox{acos}\left(+\cos({\widetilde \Delta}) + \cos(\omega) - 1 \right), \\ \hspace{0.685in} \mbox{if}  \hspace{0.05in} {\widetilde \Delta}-\pi \hspace{0.08in} \le \omega < \hspace{0.105in} 0, \\
\hspace{-0.2in} q(\omega) \cdot \mbox{acos}\left(+\cos({\widetilde \Delta}) - \cos(\omega) + 1 \right), \\ \hspace{0.69in} \mbox{if}  \hspace{0.25in} 0 \hspace{0.18in} \le \omega < \ {\widetilde \Delta}, \\
\hspace{-0.2in} q(\omega) \cdot \mbox{acos}\left(-\cos({\widetilde \Delta}) + \cos(\omega) + 1 \right), \\ \hspace{0.72in} \mbox{if}  \hspace{0.21in} {\widetilde \Delta}  \hspace{0.19in} \le  \omega  < +\pi, \\
\end{array}
\right.
\end{eqnarray}
\item If  ${\widetilde \Delta} \in [-\pi, 0)$, 
\begin{eqnarray}
\label{Oaknin_transformation_Inv}
\hspace{-0.15in}
L(\omega; {\widetilde \Delta}) =  
\left\{
\begin{array}{c}
q(\omega) \cdot \mbox{acos}\left(-\cos({\widetilde \Delta}) + \cos(\omega) + 1 \right), \\ \hspace{0.650in} \mbox{if}   \hspace{0.11in} -\pi \hspace{0.15in} \le \omega < {\widetilde \Delta}, \\
q(\omega) \cdot \mbox{acos}\left(+\cos({\widetilde \Delta}) - \cos(\omega) + 1 \right), \\ \hspace{0.66in} \mbox{if}   \hspace{0.30in} {\widetilde \Delta} \hspace{0.1in} \le \omega < \hspace{0.05in} 0, \\
q(\omega) \cdot \mbox{acos}\left(+\cos({\widetilde \Delta}) + \cos(\omega) - 1 \right), \\ \hspace{0.92in} \mbox{if}  \hspace{0.23in} 0 \hspace{0.22in} \le \omega < {\widetilde \Delta} +\pi, \\
q(\omega) \cdot \mbox{acos}\left(-\cos({\widetilde \Delta}) - \cos(\omega) - 1 \right), \\ \hspace{0.75in} \mbox{if}  \hspace{0.16in} {\widetilde \Delta} +\pi \hspace{0.00in} \le \omega < +\pi, \\
\end{array}
\right.
\end{eqnarray}
\end{itemize}
where  
\begin{eqnarray*}
q(\omega) = \mbox{sign}((\omega - {\widetilde \Delta}) \mbox{mod} ([-\pi, \pi))),
\end{eqnarray*}
and the function $y=\mbox{acos}(x)$ is defined in its main branch, such that $y \in [0, \pi]$ while $x \in [-1, +1]$. The transformation laws (\ref{Oaknin_transformation}), (\ref{Oaknin_transformation_Inv}) define a commutative group. \\
\\
Accordingly, we define the probability of each one of these possible configurations to happen as:
\begin{eqnarray}
\label{free-will}
|\mbox{d}\omega_B \ g(\omega_B)| = |\mbox{d}\omega_A \ g(\omega_A)|, 
\end{eqnarray}
where 
\begin{equation}
\label{density_distribution}
g\left(\omega\right) = \frac{1}{4} \left|\sin\left(\omega\right)\right|.
\end{equation}  
Equation (\ref{free-will}) states that the probability density distribution (\ref{density_distribution}) is functionally invariant under the transformation laws (\ref{Oaknin_transformation}), (\ref{Oaknin_transformation_Inv}), as required by symmetry considerations. More importantly, equation (\ref{free-will}) states also that the probability to occur of any of the possible configurations is invariant under these coordinate transformations or, in other words, that our model complies with the 'free will' assumption. \\
\\

\begin{figure}
\centering
\vspace{-1.0in}
\includegraphics[height=9cm]{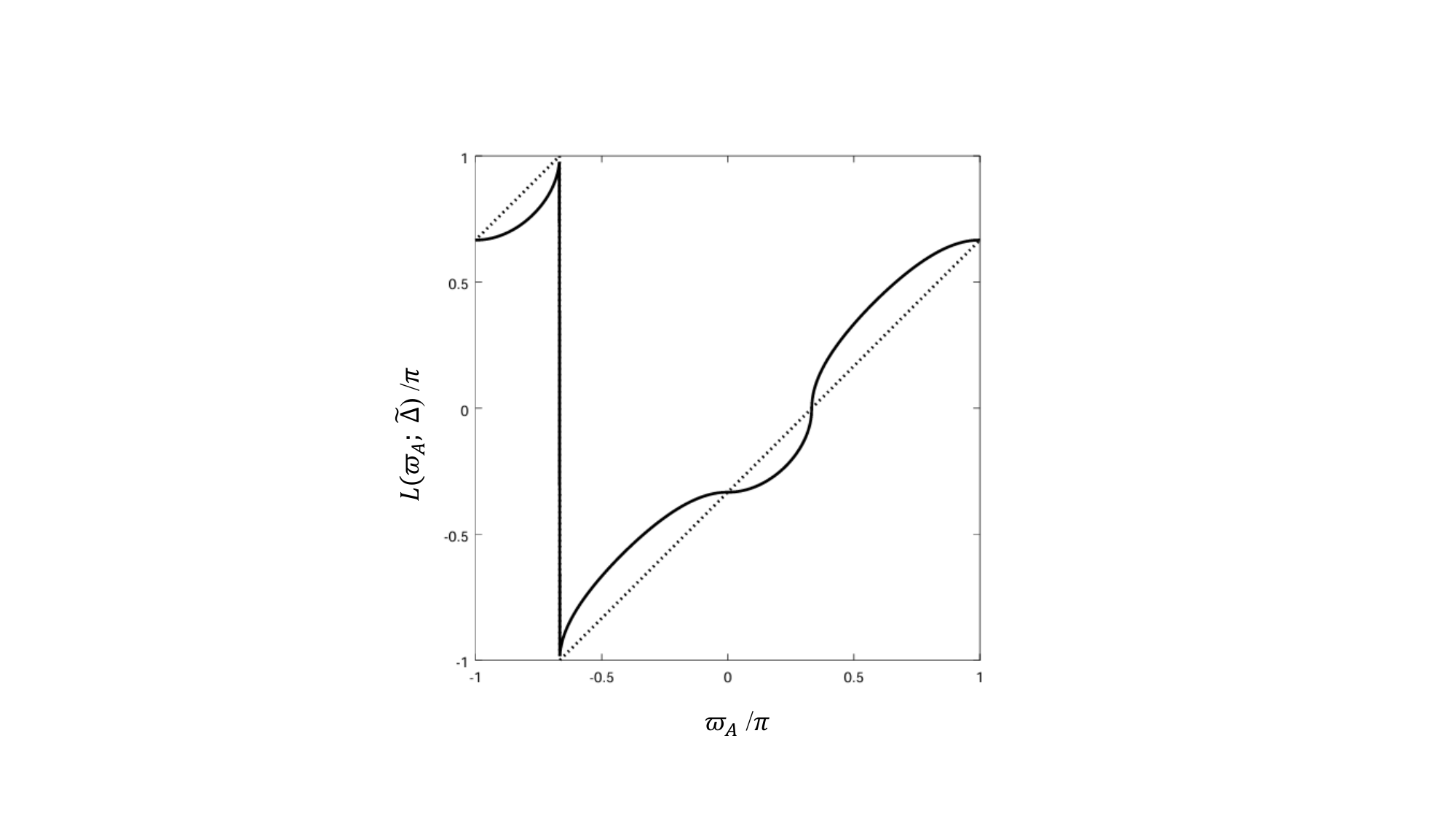}
\fontsize{10}{0}
\fontsize{10}{0}
\caption{Plot of the transformation $L(\omega; \Delta), \ \omega \in [-\pi,\pi)$, with ${\widetilde \Delta}=\pi/3$ (solid line), compared to the corresponding linear transformation (dotted line).}
\end{figure}

Then, using the transformation laws (\ref{Oaknin_transformation}), (\ref{Oaknin_transformation_Inv}) that relate $\omega_A$, $\omega_B$ and the angle ${\widetilde \Delta}$, it is straightforward to check that,
\\

\begin{eqnarray*}
\label{four_subsets}
(S_A=+1; S_B=+1) & \Longleftrightarrow & \omega_A \in [0, {\widetilde \Delta}) \\
(S_A=+1; S_B=-1) & \Longleftrightarrow & \omega_A \in [{\widetilde \Delta}, \pi) \\
(S_A=-1; S_B=+1) & \Longleftrightarrow & \omega_A \in [{\widetilde \Delta}-\pi, 0) \\
(S_A=-1; S_B=-1) & \Longleftrightarrow & \omega_A \in [-\pi, {\widetilde \Delta}-\pi),
\end{eqnarray*}
whose probabilities to happen are: 
\begin{eqnarray*}
p\left(+1,+1\right) & =  & \int_0^{{\widetilde \Delta}} g(\omega_A) \ d\omega_A \hspace{0.15in} = \ \frac{1}{4}\left(1 - \cos ({\widetilde \Delta})\right), \\
p\left(+1,-1\right)  & =  & \int_{{\widetilde \Delta}}^{\pi} g(\omega_A) \ d\omega_A \hspace{0.21in} = \ \frac{1}{4}\left(1 + \cos ({\widetilde \Delta})\right), \\
p\left(-1, +1\right) & = & \int_{{\widetilde \Delta}-\pi}^{0} g(\omega_A) \ d\omega_A \hspace{0.06in} = \ \frac{1}{4}\left(1 + \cos ({\widetilde \Delta})\right), \\
p\left(-1,-1\right) & = & \int_{-\pi}^{{\widetilde \Delta}-\pi} g(\omega_A) \ d\omega_A = \ \frac{1}{4}\left(1 - \cos ({\widetilde \Delta})\right).
\end{eqnarray*}
Hence, the statistical correlation $E({\widetilde \Delta})$ between the outcomes of the two polarization measurements is given by:
\begin{eqnarray*}
E(\Delta) = p(+1,+1) + p(-1,-1) - p(+1,-1) - p(-1,+1) = -\cos\left(\Delta\right),
\end{eqnarray*}
while the average outcome of each of them is zero independent of their setting, which, as advanced above, reproduces the predictions of Quantum Mechanics. \\
\\

It is straightforward to notice that Bell's theorem does not hold for this model because of  the non-linear transformation laws (\ref{Oaknin_transformation},\ref{Oaknin_transformation_Inv}) that relate the angular orientations of the pair of qubits in the frames of reference defined by the two measurement devices. Bell's theorem - in its original formulation - states that in generic models with locally defined elements of reality the following constraint on the statistical correlations between the outcomes of the two detectors must hold for any pair of values ${\widetilde \Delta}_1$, ${\widetilde \Delta}_2$ of the relative orientation between them:
\\

\begin{equation}
\label{Bell}
\left| E({\widetilde \Delta}_2) - E({\widetilde \Delta}_1) \right| \le 1 + E({\widetilde \Delta}_2 - {\widetilde \Delta}_1).
\end{equation}
On the other hand, in the model discussed here
\\

\begin{eqnarray}
\nonumber
\label{Bell_inequality_mebutal}
\left| E({\widetilde \Delta}_2) - E({\widetilde \Delta}_1) \right| = 
 4 \int_{{\widetilde \Delta}_1}^{{\widetilde \Delta}_2} d\omega_A \ g(\omega_A) = 1 + \left(4 \int_{{\widetilde \Delta}_1}^{{\widetilde \Delta}_2} d\omega_A \ g(\omega_A) -1\right),
\end{eqnarray}
where we have assumed without any loss of generality that $\pi \ge {\widetilde \Delta}_2 \ge {\widetilde \Delta}_1 \ge 0$. Since
\\

\begin{eqnarray*}
4 \int_{{\widetilde \Delta}_1}^{{\widetilde \Delta}_2} d\omega_A \ g(\omega_A) -1 = 4 \int_0^{\mbox{\small{arc-cos}}(\cos({\widetilde \Delta}_2)-\cos({\widetilde \Delta}_1)+1)} d\omega'_A \ g(\omega'_A) - 1
\end{eqnarray*}
is not necessarily smaller than
\\

\begin{equation}
\label{Bell_inequality_mebutal2}
E({\widetilde \Delta}_2 - {\widetilde \Delta}_1) = 4 \int_0^{{\widetilde \Delta}_2 - {\widetilde \Delta}_1} \ d\omega'_A \ g(\omega'_A) - 1,
\end{equation}
inequality (\ref{Bell}) does not hold for our model.
\\

The ultimate reason why the model is not constrained by Bell's theorem is the fact that these transformation laws (\ref{Oaknin_transformation},\ref{Oaknin_transformation_Inv}) allow for a non-zero geometric phase ${\widetilde \alpha} \neq 0$ to appear over a closed sequence of ccordinate transformations:
\\

\begin{equation}
\omega \ \ \xrightarrow{{\widetilde \Delta}_1} \ \ w'\equiv L(w; {\widetilde \Delta}_1) \ \ \xrightarrow{{\widetilde \Delta}_2} \ \  \omega''\equiv L(\omega'; {\widetilde \Delta}_2)  \xrightarrow{-({\widetilde \Delta}_1 + {\widetilde \Delta}_2)} \ \ \omega'''\equiv L(\omega''; -({\widetilde \Delta}_1 + {\widetilde \Delta}_2)) = L(\omega; {\widetilde \alpha}).
\end{equation}

\section{Summary}
\vspace{0.2in}

We have presented a statistical model built upon locally defined {\it elements of reality} that fully reproduces the predictions of Quantum Mechanics for the Bell experiment. Furthermore, the model explicitly complies with the 'free will' requirement. The crux of the model is the observation that a non-zero geometric phase may appear in the description of the pair of qubits with respect to closed sequences of settings of the measurement devices, thus rendering Bell's theorem inapplicable. Our model thus completes the quantum mechanical description of Bell's states in the terms envisioned by Einstein, Podolsky and Rosen.

\end{document}